\documentclass[british,english]{article}
\usepackage[T1]{fontenc}
\usepackage[latin9]{inputenc}
\usepackage{babel}
\usepackage{float}
\usepackage{amsmath}
\usepackage{amssymb}
\usepackage{graphicx}
\usepackage[unicode=true,
 bookmarks=true,bookmarksnumbered=false,bookmarksopen=false,
 breaklinks=false,pdfborder={0 0 1},backref=false,colorlinks=false]
 {hyperref}
\hypersetup{pdftitle={\@shorttitle},
 pdfauthor={\@shortauthor},
 pdfsubject={\@Subject},
 pdfkeywords={Computer Graphics Forum, EUROGRAPHICS},
   colorlinks,linkcolor=blue,citecolor=blue,urlcolor=blue,    bookmarks=false,    pdfpagemode=UseNone}

\makeatletter

\providecommand{\tabularnewline}{\\}

\usepackage{tikz,tkz-tab,tkz-graph}
\makeatother

\begin{document}

\title{Networks of Collaborations:\\
Hypergraph Modeling and Visualisation}

\author{Xavier $\mbox{Ouvrard}^{1,2}$\qquad{}Jean-Marie $\mbox{Le Goff}^{1}$\qquad{}Stéphane
Marchand-$\text{Maillet}^{2}$\\
\\
1~:~CERN\qquad{}2~:~University of Geneva\\
{\small{}\{xavier.ouvrard\}@cern.ch}}
\maketitle
\begin{abstract}
The acknowledged model for networks of collaborations is the hypergraph
model. Nonetheless when it comes to be visualized hypergraphs are
transformed into simple graphs. Very often, the transformation is
made by clique expansion of the hyperedges resulting in a loss of
information for the user and in artificially more complex graphs due
to the high number of edges represented. The extra-node representation
gives substantial improvement in the visualisation of hypergraphs
and in the retrieval of information. This paper aims at showing qualitatively
and quantitatively how the extra-node representation can improve the
visualisation of hypergraphs without loss of information.
\end{abstract}

\section{Introduction}

Euler in 1736, with the Seven Bridges of Königsberg, was the first
one to write a paper on a graph related problem. The word graph itself
was introduced by Sylvester in 1878. A lot has been done since these
days in particular theoretical developments during the first half
of the twentieth century. With the increasing calculation power of
computers, graphs have now taken an important practical place. The
recent emergence of social networks as a means of retrieving information
from data has boosted the use of graphs and hypergraphs. The rising
of the Big Data era with its huge amount of data calls for powerful
analytical and visualising tools. In addition to their modelisation
and the identification of particular features, the study of such networks,
including collaboration networks, has to address the retrieval of
important information that can enrich the visual perception of the
dataset.

The graph theory - the result of an extensive study of graphs over
the years - provides the foundations for graph modeling. The first
famous model was the random graph developed by Erdös and Renyi in
1959. It was followed by many others, such as the small-world model
of Watts and Strogatz at the end of the 20th century, shown as being
an illustration of the six handshakes lemma found by Karinthy in 1929
that stipulates that two persons in the world are no further away
from any other people than six handshakes. In 2001, the work of Watts
and Strogatz was enhanced by Albert and Barabasi in the modeling of
scale-free networks.

At this point, the specifities of collaboration networks must be stressed.
\cite{NEWMAN 2001} explains that collaborators of a publication have
a $m$-adic relationship - in the sense they are attached to the publication
-, where $m$ is the number of collaborators in a collaboration. Nonetheless,
this $m$-adic relationship is approximated by a 2-adic relationship
in between pairs of collaborators when it comes to be studied. The
same approximation is made in many other studies such as \cite{RAMASCO 2004}.

The reasons for this approximation are numerous. It enables the use
of classical graphs techniques and properties when studying the different
characteristics of collaboration networks, such as degree distribution,
clustering coefficient, and when applying quantifying metrics. Today,
many different techniques helping the retrieval of information from
graphs are available. Amongst them, clustering techniques play an
important role since they facilitate the extraction of information
from networks. An efficient analysis algorithm that can be run on
graphs is the Louvain's algorithm introduced in \cite{BLONDEL =000026 al. 2008},
which is strongly based on the graph structure of the dataset; in
this kind of algorithm the main issue is to give meaning to the resulting
clusters.

This 2-adic relationship approximation has been developped in many
articles, where even if the $m-$adic relationship of the data was
pointed to be more pertinent, this $m$-adic relationship was not
used when getting to clustering. Since the end of the 2000s years,
the limitations of the 2-adic approach is more and more challenged,
as it leads to a partial loss of information contained in the $m$-adic
relationship. As a result, in \cite{ESTRADA 2005} the authors modelize
complex networks by hypergraphs. 

\cite{BERGE 1973} introduced hypergraphs as a means to generalize
the graph approach. Hypergraphs preserve the $m$-adic relationship
becoming the natural modeling of collaboration networks. An hypergraph
$\mathcal{H}=\left(V,H\right)$ on a finite set of vertices (or nodes)
$V=\left\{ x_{1}\,;\,x_{2};\,...\,;\,x_{n}\right\} $ is defined as
a family of hyperedges $H=\left(E_{1},E_{2},...,E_{m}\right)$ where
each hyperedge is a non-empty subset of $V$ and such that $\bigcup\limits _{i=1}^{m}E_{i}=V$.
This means that in an hypergraph, an hyperedge links one or more vertices.
In \cite{BRETTO 2013}, this last hypothesis is relaxed to enable
isolated vertices in hypergraphs, opening the use of hypergraphs in
various collaboration networks. Actually, an hypergraph can also be
seen as a set of sets.

Hypergraphs features are very similar to those of graphs with some
arrangements to account for their differences in structure.

The \textbf{order} of the hypergraph is defined as $\left|V\right|$. 

The \textbf{rank} of an hypergraph is the maximum of the cardinalities
of the hyperedges while the \textbf{anti-rank} corresponds to the
minimum. An hypergraph is said \textbf{simple} if there's no multiple
hyperedges in between a set of nodes.

The \textbf{node degree} corresponds to the number of hyperedges that
the node participates in. It is also designated as \textbf{hyperdegree}
in some articles. The \textbf{distance in between two nodes} is the
minimal number of hyperedges that connect the two nodes.

The \textbf{incidence matrix} $E$ of an hypergraph is the matrix
whose rows represent the nodes $x_{1},...,x_{n}$ and whose columns
represent the hyperedges $E_{1},...,E_{m}$ and where the coefficient
$a_{ij}=1$ when $x_{i}\in E_{j}$, and $a_{ij}=0$ when $x_{i}\notin E_{j}$.

The \textbf{adjacency matrix} $A$ of an hypergraph is a square matrix
whose lines and rows represent the nodes $x_{1},...,x_{n}$ and where
the coefficient $a_{ij}$ is the number of hyperedges where $x_{i}$
and $x_{j}$ are present together.

\cite{ESTRADA 2005} introduces particular features to characterize
hypergraphs. The authors define the relationship between the adjacency
and the incidence matrix as: $A=EE^{T}-D$ where $D$ is the diagonal
matrix containing vertex degrees. They evaluate the \textbf{centrality
of a node} in a simple hypergraph, by orthogonalizing the adjacency
matrix in $A=UDU^{T}$, where $U=\left(u_{ij}\right)$ $D=diag\left(\lambda_{1},...,\lambda_{n}\right)$
is the diagonal matrix formed of the eigenvalues $\lambda_{i}$ ($1\leqslant i\leqslant n$)
of $A$.

The sub-hypergraph centrality is defined as the sum of the closed
walks of different lengths in the network, starting and ending at
a given vertex.

The sub-hypergraph centrality $C_{SH}(i)$ can be calculated in a
simple hypergraph as: $C_{SH}(i)=\sum\limits _{j=1}^{n}\left(u_{ij}\right)^{2}\mbox{e}^{\lambda_{j}}$. 

They also define a \textbf{clustering coefficient} for an hypergraph
as: $C(H)=\dfrac{6\times\mbox{number of hyper-triangles}}{\mbox{number of 2-paths}}$
where a hyper-triangle is defined as a sequence of three different
vertices that are separated by three different hyperedges $v_{i}E_{p}v_{j}E_{q}v_{k}E_{r}v_{i}$
and a 2-path is a sequence $v_{i}E_{p}v_{j}E_{q}v_{k}.$

In \cite{TARAMASCO 2010}, the authors study the academic team formation
using epistemic hypergraphs where hyperedges are subsets of unions
of a set of agents and a set of concepts. They introduce new features
to characterize the evolution of collaboration networks taking into
account the hypergraphic nature of networks. This paper brings keystones
in the study of a bidimensionnal hypergraph and show how the keeping
of $m$-adic relationships can help to gain in the understanding of
the evolution of a network.

This paper aims at showing that network of collaborations have an
efficient modelisation by hypergraphs and those hypergraphs have a
suitable visualisation that enrich the data visualisation experience.
It is to our knowledge the first time such an experimental comparison
is made for hypergraphs visualisation. Section \ref{sec:Hypergraphs-of-Collaboration}
a theoretical framework of collaboration networks viewed as hypergraphs.
Section \ref{sec:Visualisation-of-Hypergraphs} provides a survey
of the different representations of hypergraphs that can be done,
the need of evaluation of such representations is pointed out. Finally
Section \ref{sec:Experimental-Approach} shows experimentally that
efficient representation of such hypergraphs can be made to have a
valuable visualisation that enhance the understanding of the underlying
data.

\section{Hypergraphs of Collaboration Networks\label{sec:Hypergraphs-of-Collaboration}}

Different information such as authors and their affiliation to organisations
can be retrieved from scientific publication metadata. Studying the
relationships between authoring organisations helps having a better
understanding of the world of science. This section aims at giving
a theoretical framework of scientific collaboration networks.

Let consider that in a paper $p$, there are $o_{p}$ organisations
and $c_{p}$ countries where those organisations are seated. This
group of organisations can be viewed as a set $O_{p}=\left\{ \omega_{1}\,;\,...\,;\,\omega_{o_{p}}\right\} $
which is attached to paper $p$ as well as $C_{p}=\left\{ \gamma_{1},...,\gamma_{c_{p}}\right\} $
is the set of countries attached to $p$.

Another revealing information contained in a publication is the keywords
put by the authors, mentionned as author keywords. With the same approach,
let $K_{p}=\left\{ \xi_{1}\,;...\,;\xi_{k_{p}}\right\} $ be the set
of the $k_{p}$ author keywords found in publication $p$. $O_{p}$,
$C_{p}$, $K_{p}$ constitute amongst other relevant information a
multi-set of attributes from different dimensions found in publication
$p$. Performing a semantic search on a datastore for a particular
topic will return a set of $s$ publications $S=\left\{ p_{1},...,p_{s}\right\} $
each having sets of authoring organisations, countries and keywords.

Some attributes are common to all or a part of the publications resulting
from this search. For instance, from this set of papers, the set of
cited organisations can be defined as $O_{S}=\bigcup\limits _{p\in S}O_{p}$.
This set will form the set of nodes that will be represented in an
hypergraph. The set of collaborations $O_{p}$ extracted from the
paper $p$ can be viewed as an hyperedge. The same approach can be
taken for countries or keywords, building two other uni-dimensional
hypergraphs.

Of course this approach is transferable to any other relevant kind
of attributes in an article such as city, journal categories,... Let
$\alpha$ be an attribute type that can be found in paper $p$, in
$a_{p}$ quantity. The set of attributes of type $\alpha$ attached
to this paper is $A_{\alpha,p}=\left\{ \alpha_{1},...,\alpha_{a_{p}}\right\} $.
$A_{p}$ is the set of co-attributes instances. 

As a consequence, attributes of type $\alpha$ that are common to
two papers $p_{1}$ and $p_{2}$, are $A_{\alpha,p_{1}}\cap A_{\alpha,p_{2}}$.

If a search $S$ is performed, then the set of values for the attributes
of type $\alpha$ in the results returned is $A_{\alpha,S}=\bigcup\limits _{p\in S}A_{\alpha,p}$.
As there will be one set $A_{\alpha,p}$ per paper, possibly empty,
the search result for this attributes of type $\alpha$ is a set of
sets, written:
\[
\mathcal{A}_{\alpha,S}=\left\{ A_{\alpha,p}|p\in S\right\} .
\]

$A_{\alpha,S}$ can be viewed as a set of nodes of attributes of type
$\alpha$ and $\mathcal{A}_{\alpha,S}$ as a set of hyperedges of
coattributes of type $\alpha$.

The hypergraph 
\[
\mathcal{H}_{\alpha,S}=\left(A_{\alpha,S},\,\mathcal{A}_{\alpha,S}\right)
\]
 is the hypergraph of co-$\alpha$ type attributes in the search $S$.
It is a representation of the collaborations of co-$\alpha$ type
attributes that are included in papers from search $S$. There are
as many hyperedges as the number of papers in which the team is involved.
Also it can be valuable to set a weight to each different hyperedge
with an initial value of 1. If there exist two hyperedges that are
identical - that is two papers having the same set of attributes attached
to it - then they can be merged into one hyperedge with a weight that
is the sum of the weights of the two initial hyperedges. And the final
hypergraph is a pondered hypergraph of teams of co-$\alpha$ type
attributes in the search S.

If an hypergraph corresponding to an other type $\alpha'$ is needed
then the attributes of this type in the search S will lead to hypergraph
$\mathcal{H}_{\alpha',S}=\left(A_{\alpha',S},\,\mathcal{A}_{\alpha',S}\right).$
Therefore, one can build a similar hypergraph for each attribute type
in a publication metadata instance and create a set of typed hypergraphs
each representing individual views on the multi-dimensional network
$S$ and connected via subsets of paper metadata instances in $S$.

This approach can be generalized to other datasets such as patents,
allowing to view values of an attribute with a type common to different
datasets into a single hypergraph. For instance, the hypergraph of
authoring organisations for a search on patents and publications related
a particular topic. 

Ultimately, by building a multi-dimensional network organised around
attribute types, one can retrieve very valuable information from combined
data sources. This process can be extended to any number of data sources
as long as they share one or more attribute types. If this is not
the case, we will have unconnected networks that cannot be navigated
accross.

\section{Visualisation of Hypergraphs\label{sec:Visualisation-of-Hypergraphs}}

Visualizing hypergraphs is an issue that can prevent their intensive
use. In this Section a survey of existing hypergraph's representations
in litterature is done pointing out the issues raised and showing
the need of comparison of the two main representations.

\cite{ESTRADA 2005} in an in-depth paper on hypergraphs have skimmed
the subject of visualisation with only one kind of representation:
the Venn's diagram - a usual representation for sets - is relevant
for small hypergraphs, but will be hard to use for large hypergraphs.

Based on the work of \cite{M=0000C4KINEN 1990}, \cite{JUNGHANS 2008}
classifies the hypergraphs' visualisations as the edge standard that
makes connections between nodes of an hyperedge and the subset standard
that makes closed curves encompassing these nodes. The Venn\textquoteright s
diagram is part of the subset standard.

\begin{figure}[H]
\begin{center}%
\begin{tabular}{cc}
\newcount\tempcount
\def \n {9}
\def \mn {8}
\def \radius {1cm}
\begin{tikzpicture}[transform shape, every node/.style = {scale = 0.5}]
\foreach \s in {1,...,\n} {   
     \node[draw, circle, fill=black] (N-\s) at ({360/\n * (\s - 1)}:\radius) {};
}    
\foreach \ni in {1,...,\mn}{
   \tempcount=\ni
   \advance\tempcount by 1
   \foreach \nj in {\the\tempcount,...,\n}{
      \path (N-\ni) edge (N-\nj);
   }
}
\end{tikzpicture} & \newcount\tempcount
\def \n {9}
\def \mn {8}
\def \radius {1cm}
\begin{tikzpicture}[transform shape, every node/.style = {scale = 0.5}]
\node[draw,circle,fill=white,scale=0.1] (N-0) at (0,0) {};
\foreach \s in {1,...,\n} {   
     \node[draw, circle, fill=black] (N-\s) at ({360/\n * (\s - 1)}:\radius) {};
}    
\foreach \ni in {1,...,\n}{
   \path (N-0) edge (N-\ni);
}
\end{tikzpicture}\tabularnewline
\end{tabular}\end{center}

\caption{Clique view vs extra-node view of an Hyperedge}
\label{Fig : Clique view vs Extra node view of an Hyperedge}
\end{figure}
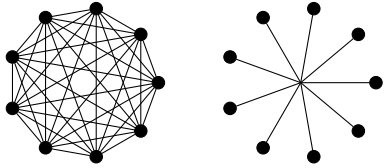

In the edge standard, there are two main representations: the clique
expansion and the extra- (or crux-) node representation. In the clique
expansion, each node of an hyperedge is connected with all other nodes
by an edge. Therefore an hyperedge of size $n$ is represented by
$\dfrac{n(n-1)}{2}$ edges. In the extra-node representation, only
$n$ edges are needed. Though the potential gain is $\dfrac{n(n-3)}{2}$
and is strictly positive above 3. These two views of one hyperedge
are illustrated in Figure \ref{Fig : Clique view vs Extra node view of an Hyperedge}. 

\vspace{-0.5cm}

Nonetheless the gain in edges is only a potential gain. Some unfavorable
case for extra-node representation can be easily exhibited as Figure
\ref{Fig : Extra node unfavorable cases} shows it. This implies to
study carefully the distribution of hyperedges and the way they intersect.
It outlines the need of deeper understanding of the pertinence of
such an approach.

\begin{figure}[H]
\resizebox{\columnwidth}{!}{%
\begin{tabular}{|c|c|}
\hline 
Clique view & extra-node view\tabularnewline
\hline 
\newcount\tempcount
\def \n {5}
\def \mn {4}
\def \radius {1cm}
\begin{tikzpicture}[transform shape, every node/.style = {scale = 0.5}]
\foreach \s in {1,...,\n} {   
     \node[draw, circle, fill=black] (N-\s) at ({360/\n * (\s - 1)}:\radius) {};
}    
\foreach \ni in {1,...,\mn}{
   \tempcount=\ni
   \advance\tempcount by 1
   \foreach \nj in {\the\tempcount,...,\n}{
      \path (N-\ni) edge (N-\nj);
   }
}
\end{tikzpicture} & \newcount\tempcount
\def \n {5}
\def \mn {4}
\def \radius {1cm}
\begin{tikzpicture}[transform shape, every node/.style = {scale = 0.5}]
\node[draw,circle,fill=white,scale=0.1,color=red] (N-0) at (0,0) {};
\foreach \s in {1,...,\n} {   
     \node[draw, circle, fill=black] (N-\s) at ({360/\n * (\s - 1)}:\radius) {};
}    
\foreach \ni in {1,...,\n}{
   \path[color=red] (N-0) edge (N-\ni);
}
\node[draw,circle,fill=white,scale=0.1,color=green] (N-6) at (0.5,0.25) {};
\foreach \ni in {1,...,3}{
   \path[color=green] (N-6) edge (N-\ni);
}
\node[draw,circle,fill=white,scale=0.1,color=blue] (N-7) at (-0.5,-0.25) {};
\foreach \ni in {3,...,5}{
   \path[color=blue] (N-7) edge (N-\ni);
}
\end{tikzpicture}\tabularnewline
\hline 
In this case: 10 edges, 5 nodes & In this case: 11 edges, 5 nodes and 3 extra-nodes\tabularnewline
\hline 
\end{tabular}

}

\caption{Unfavorable case for extra-node view}
\label{Fig : Extra node unfavorable cases}
\end{figure}

\vspace{-0.5cm}

\cite{JUNGHANS 2008} focuses on the drawing of hyperedges so that
they don't intersect cluster groups, giving a solution based on force
attraction/repulsion drawing of hyperedges. The author provides an
interesting synthesis on the cognitive load of such representation.
Nonetheless no systematic comparison between the clique and extra-node
representation is made.

Some other hypergraphs' representations exist such as the pie-chart
node approach presented in \cite{PAQUETTE 2011}, which is relevant
for hypergraphs when hyperedges are not too intersecting one another.
There is also the radial approach presented in \cite{KEREN =000026 al. 2013},
which is surely valuable in the case of small order hypergraphs, but
will be hard to implement for large hypergraphs. Some other techniques
derived from set representations could be of interest. The interesting
reader can refer to \cite{ALSALLAKH =000026 al. 2016}.

The rest of this study will focus on the two main views of an hypergraph:
the clique view and the extra-node view. One can debate on the pertinence
of each to represent collaboration networks. In the clique view, collaborations
are seen as 2-adic interactions and the information on the meso-structure
is lost, as shown in \cite{TARAMASCO 2010}. In the extra-node view
the $m$-adic interactions preserve the information on individual
collaborations. Keeping this $m$-adic relationship is interesting
for many reasons. 

\section{Experimental Approach\label{sec:Experimental-Approach}}

Prevalently, people and organisations are not working alone, they
collaborate in teams that appear as co-authors in publications. The
extra-node view preserves this information since different individual
collaborations are represented as separate hyperedges enabling a direct
visualisation of the contribution of individual collaborators into
different collaborations which is clearly not possible with the clique
view as shown on the Figure \ref{Fig : Extra node unfavorable cases}.
Furthermore, large collaborations tend to be over-emphasized in the
clique view which artificially enhances the visual perception of the
relative importance of these collaborations in their respective clusters
to the detriment of smaller collaborations with more activity (more
co-publications, with less collaborators). 

All these arguments show the importance of a detailed study, including
statistics on potential gain and achieved gain as well as quantitative
and qualitative comparison of the clique and extra-node view. This
is presented in this section.

The study is conducted on a large collection of publications \textendash{}
about 45 M records of metadata instances - that has been processed
to build a multidimensional network and store it in \cite{Neo4j}
a graph database. Cypher - the associated query language with Neo4j
- is used to retrieve the values of vertex attributes of two types:
organisation and keyword.

The first part of the study addresses the potential gain in edges
with respect to all the organisations and authors keywords collaborations
contained in the database in order to obtain a maximum value for this
gain. This work has been done using 63 different semantic searches
on topics of importance for particle physics. The second part evaluates
quantitatively and qualitatively the gain obtained when toggling between
the clique and the extra-node views. It aims to show that an enhanced
visual perception is obtained through the extra-node view.

A laboratory environment has been developed to process all the organisation
and keyword collaborations found in publications from individual search
results and store them as hypergraphs that are then used to retrieve
statistics and build both views. Before completing the rendering of
individual views the Louvain clustering algorithm based on the work
of \cite{BLONDEL =000026 al. 2008} is performed on collaborations
followed by a cluster visual positioning algorithm, ForceAtlas2 as
exposed in \cite{JACOMY 2014} that calculates the coordinates of
clusters using an energy-based mechanism. For facilitating comparisons
between the two views of the same hypergraph, either ForceAtlas2 was
performed on the clique view and transposed to the extra-node view
or vice-versa.

\subsection{Statistical approach}

\begin{figure}[h]
\begin{centering}
\includegraphics[scale=0.4]{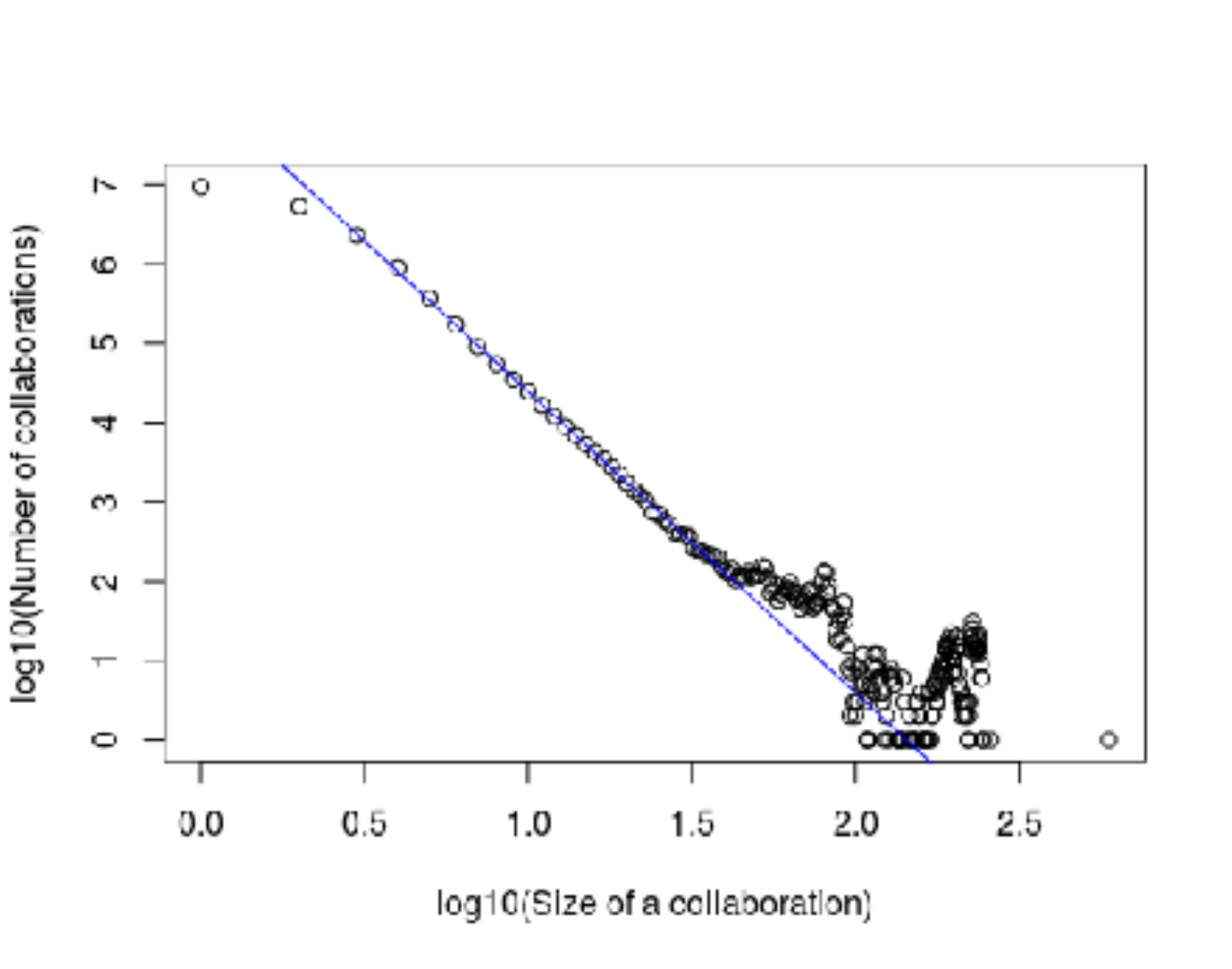}
\par\end{centering}
\centering{}\caption{$\log_{10}(\mbox{\ensuremath{\left|collaboration\right|}) vs }\log_{10}(\mbox{size of coll.)}$}
\label{Fig : n collaborations vs size of collaborations-1}
\end{figure}

The distribution of collaborations for organisations amongst the collection
of publication metadata records is shown in Figure \ref{Fig : n collaborations vs size of collaborations-1}
using log scales. In the linear part, the number of collaborations
$N$ with size $\left|C\right|$ is given by: $N\approx10^{8.199}\times\left|C\right|^{-3.799}$
with a correlation coefficient $r^{2}$ of 0.9985. The average collaboration
is 1.95.

\begin{figure}[h]
\begin{centering}
\includegraphics[scale=0.4]{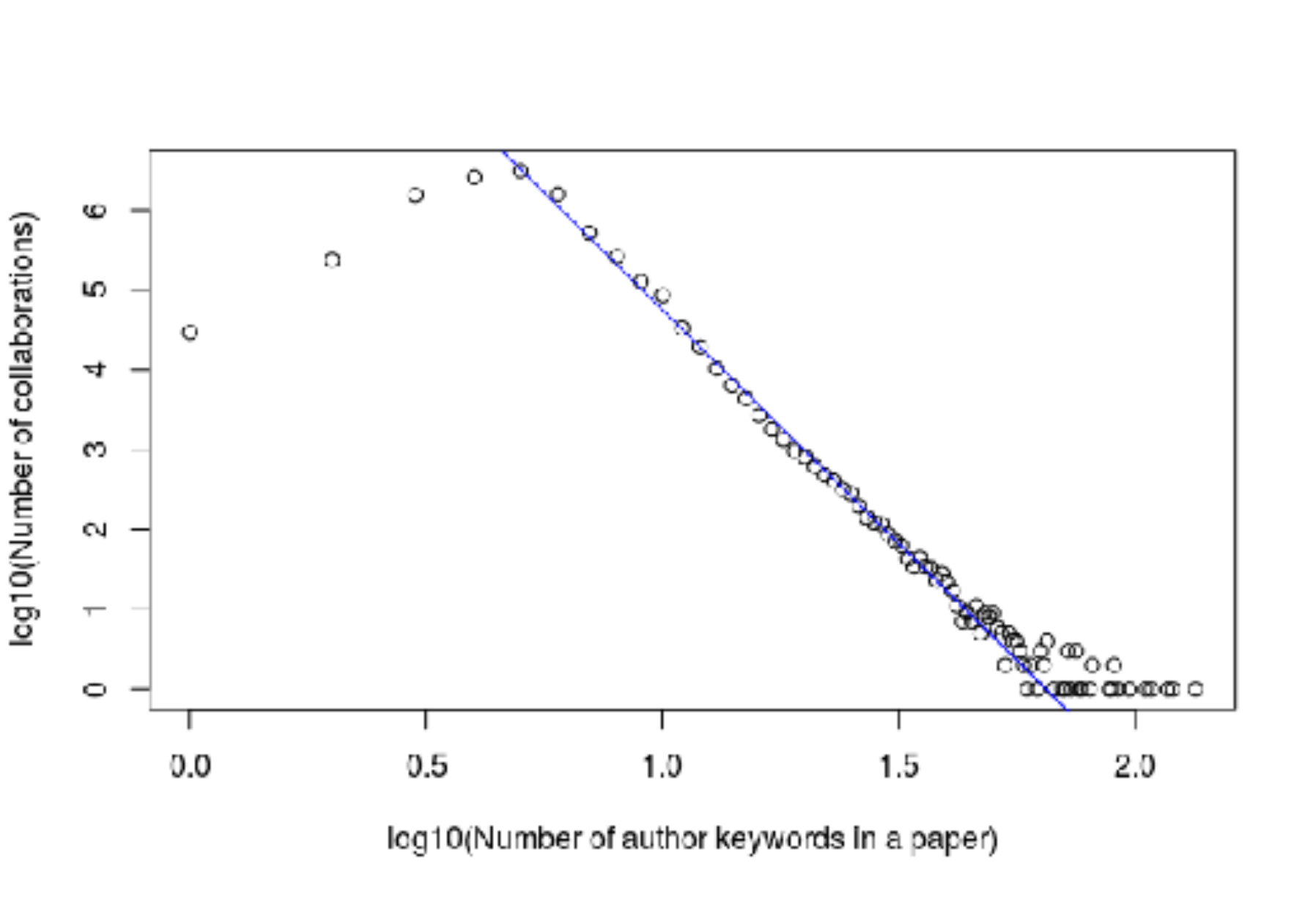}
\par\end{centering}
\centering{}\caption{$\log_{10}(\mbox{\ensuremath{\left|collaboration\right|}) vs }\log_{10}(\mbox{\ensuremath{\left|keywords\right|})}$}
\label{Fig : n collaborations vs size of collaborations-akw}
\end{figure}

The distribution of collaborations for author keywords is shown in
Figure \ref{Fig : n collaborations vs size of collaborations-akw}
using log scales. In the linear part, the number of collaborations
$N$ with $\left|K\right|$ author keywords is given by: $N\approx10^{10.6}\times\left|K\right|^{-5.852}$
with a correlation coefficient $r^{2}$ of 0.9959. The average cardinality
of co-(author keywords) set is 4.86.

Table \ref{Table : Potential gain in edges organisations} shows the
number of collaborations in the collection according to their size
and the corresponding number of edges both in the clique and extra-node
view. On average, the theoretical gain in edges - as defined in Section
\ref{subsec:Gain-in-edges} - between the two views is nearly 3. 

\begin{table}[H]
\resizebox{\columnwidth}{!}{

\begin{tabular}{|c|c|c|c|c|}
\hline 
$\left|O_{p}\right|$ & $\left|\left\{ p:\left|O_{p}\right|=k\right\} \right|$ & $n_{\mbox{edge clique view}}$ & $n_{\mbox{edge extra node view}}$ & $G_{\mbox{edge}}$\tabularnewline
\hline 
$k=$1 & 9,436,821 & x & x & x\tabularnewline
\hline 
$k=$2 & 5,331,106 & 5,331,106 & 5,331,106 & 1\tabularnewline
\hline 
$k=$3 & 2,294,535 & 6,883,605 & 6,883,605 & 1\tabularnewline
\hline 
$k=$4 & 901,023 & 5,406,138 & 3,604,092  & 1.5\tabularnewline
\hline 
$k=$5 & 370,669 & 3,706,690 & 1,853,345  & 2\tabularnewline
\hline 
$6\leqslant k\leqslant10$ & 377,253 & 8,390,494 & 2,667,516  & 3.15\tabularnewline
\hline 
$11\leqslant k\leqslant15$ & 74,102 & 3,572,175 & 616,380 & 5.80\tabularnewline
\hline 
$16\leqslant k\leqslant20$ & 14,415 & 2,110,470 & 253,194 & 8.34\tabularnewline
\hline 
$21\leqslant k\leqslant50$ & 10,776 & 4,617,743 & 309,694 & 14.91\tabularnewline
\hline 
$51\leqslant k\leqslant100$ & 3,043 & 7,387,620  & 210,079 & 35.16\tabularnewline
\hline 
$k>100$ & 920 & 17,637,388  & 175,739 & 100.36\tabularnewline
\hline 
Sum & 18,814,663 & 65,043,429 & 21,904,750 & 2.97\tabularnewline
\hline 
\end{tabular}

}\newline\caption{Organisations : Potential gain in edges}
\label{Table : Potential gain in edges organisations}
\end{table}

\vspace{-0.5cm}

Table \ref{Table : Potential gain in edges author keywords} shows
the number of occurences according to the cardinal of the co-(author
keywords) set and the corresponding number of edges both in the clique
and extra-node view. On average, the theoretical gain in edges between
the two views is nearly 2.2. 

\begin{table}[H]
\resizebox{\columnwidth}{!}{

\begin{tabular}{|c|c|c|c|c|}
\hline 
$\left|K_{p}\right|$ & $\left|\left\{ p:\left|K_{p}\right|=k\right\} \right|$ & $n_{\mbox{edge clique view}}$ & $n_{\mbox{edge extra node view}}$ & $G_{\mbox{edge}}$\tabularnewline
\hline 
$k=$1 & 29,203 & x & x & x\tabularnewline
\hline 
$k=$2 & 236,099 & 236,099 & 236,099 & 1 \tabularnewline
\hline 
$k=$3 & 1,530,790 & 4,592,370 & 4,592,370 & 1 \tabularnewline
\hline 
$k=$4 & 2,568,366 & 15,410,196 & 10,273,464 & 1.5 \tabularnewline
\hline 
$k=$5 & 3,074,370 & 30,743,700 & 15,371,850 & 2 \tabularnewline
\hline 
$6\leqslant k\leqslant10$ & 2,556,805 & 50,053,697 & 17,098,753 & 2.93 \tabularnewline
\hline 
$11\leqslant k\leqslant15$ & 73,330 & 4,929,632 & 883,074 & 5.58\tabularnewline
\hline 
$16\leqslant k\leqslant20$ & 7,570 & 1,086,347 &  131,676 & 8.25\tabularnewline
\hline 
$21\leqslant k\leqslant50$ & 3,243 & 1,079,835 & 83,424 & 12.94\tabularnewline
\hline 
$51\leqslant k\leqslant100$ & 65 & 135,482 & 4,154 & 32.61 \tabularnewline
\hline 
$k>100$ & 5 & 34,075 & 584 & 58.35\tabularnewline
\hline 
Sum & 10,079,846  & 108,301,433 &  49,630,119  &  2.22\tabularnewline
\hline 
\end{tabular}

}\newline\caption{Author keywords : Potential gain in edges}
\label{Table : Potential gain in edges author keywords}
\end{table}

\vspace{-0.5cm}

This is only a potential gain since it highly depends on how the hyperedges
are intersecting one another. As a consequence, an experimental evaluation
of the gain in edges has to be performed.

\subsection{Qualitative approach}

The qualitative approach consists in generating hypergraphs of organisations
or keywords from a subset of the collections of publication metadata
records. The aim is to have a human visual comparison of the hypergraphs
visualized as clique and extra-node views and a methodology to perform
such representations. 

From the 63 searches performed, 22 gave very large data set results
leading to extremely complicated graphs that are not directly exploitable
for the comparisons between the two views. As a consequence the graphical
comparison is made over a set of 41 searches. A typical example is
given with a search on BGO - Bismuth Germanium Oxyde - crystals.

Table \ref{Tab : Stat BGO} gives statistics. The average size of
collaborations are relatively similar for organisations and author
keywords. Nonetheless the gain in edges is much greater for the representations
of co-organisations than in the co-(author keywords') ones ; this
is due to the very variable size of collaborations of organisations
while the number of author keywords is not so different from one publication
to another.

\begin{table}[H]
\resizebox{\columnwidth}{!}{%
\begin{tabular}{|c|c|c|c|c|}
\cline{2-5} 
\multicolumn{1}{c|}{} & \multicolumn{2}{c|}{Organisations} & \multicolumn{2}{c|}{Author keywords}\tabularnewline
\cline{2-5} 
\multicolumn{1}{c|}{} & Clique view & extra-node view & Clique view & extra-node view\tabularnewline
\hline 
Number of collaborations & \multicolumn{2}{c|}{169} & \multicolumn{2}{c|}{193}\tabularnewline
\hline 
Average size of collaborations & \multicolumn{2}{c|}{3.83} & \multicolumn{2}{c|}{4.48}\tabularnewline
\hline 
Number of nodes & 349 & 439 & 597 & 783\tabularnewline
\hline 
Number of edges & 2639 & 647 & 1699 & 864\tabularnewline
\hline 
Gain in edges & \multicolumn{2}{c|}{4.07} & \multicolumn{2}{c|}{1.97}\tabularnewline
\hline 
\end{tabular}

}\newline

\caption{Statistics on the search : title:((bgo AND cryst{*}) OR (bgo AND calor{*}))
abstract:((bgo AND cryst{*}) OR (bgo AND calor{*}))}

\label{Tab : Stat BGO}
\end{table}

\vspace{-0.5cm}

The benchmark program has been configured to transfer the coordinates
calculated by ForceAtlas2's algorithm from one hypergraph representation
to the other.

This transfer was made two ways: either calculation of the coordinates
on the clique view and transfer to the nodes of the extra-node view
- the extra-node is always considered as the isobarycenter of the
hyperedge - or reciprocally.

\selectlanguage{british}%
\begin{figure}[H]
\selectlanguage{english}%
\begin{center}\includegraphics[scale=0.6]{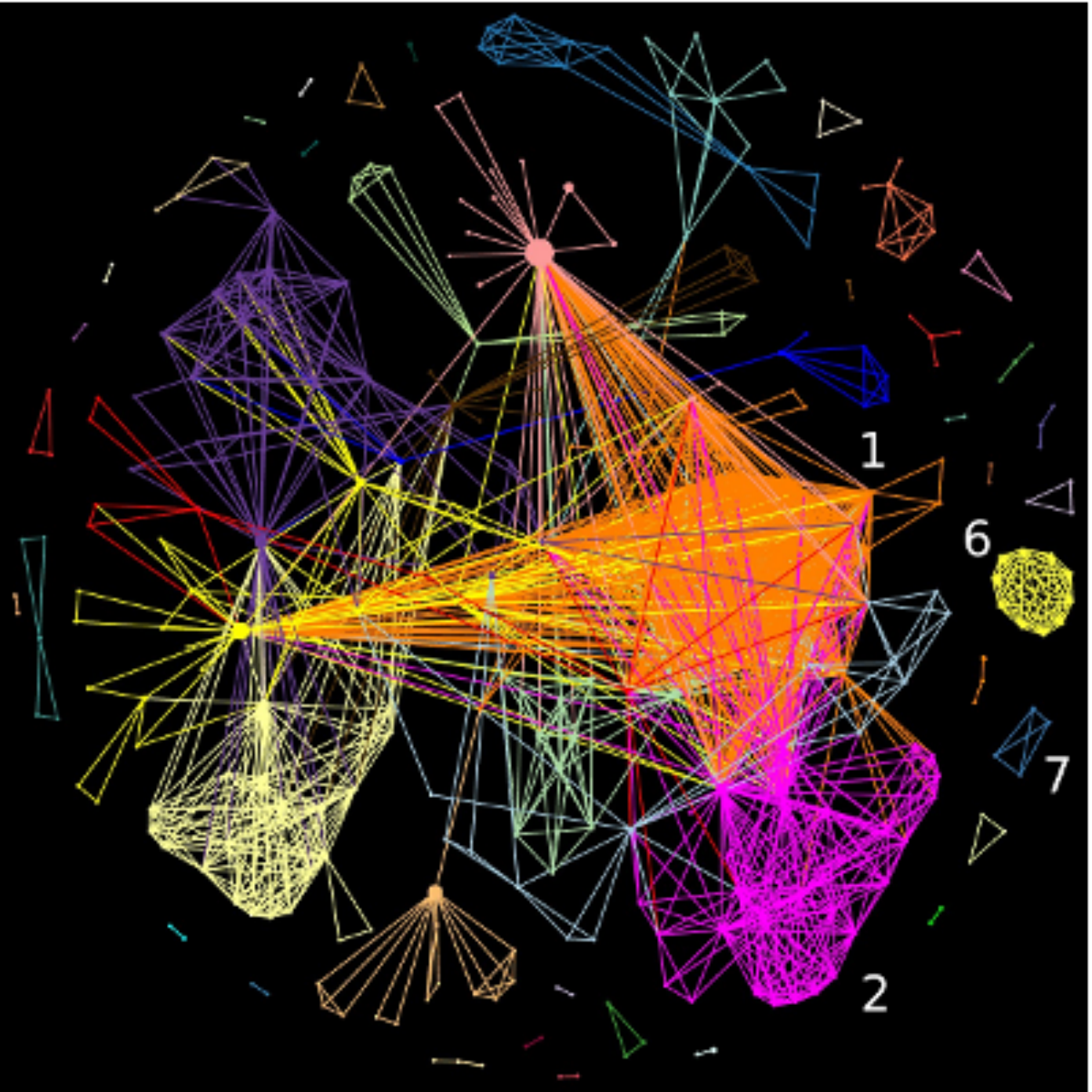}\end{center}

Sub-figure \ref{Fig exp : title:((bgo AND cryst*) OR (bgo AND calor*)) abstract:((bgo AND cryst*) OR (bgo AND calor*)) : organisations eng}
(a): Clique representation: The coordinates of nodes are calculated
by ForceAtlas2 on the extra-node view and then transfered to this
view.\newline

\begin{center}\includegraphics[scale=0.6]{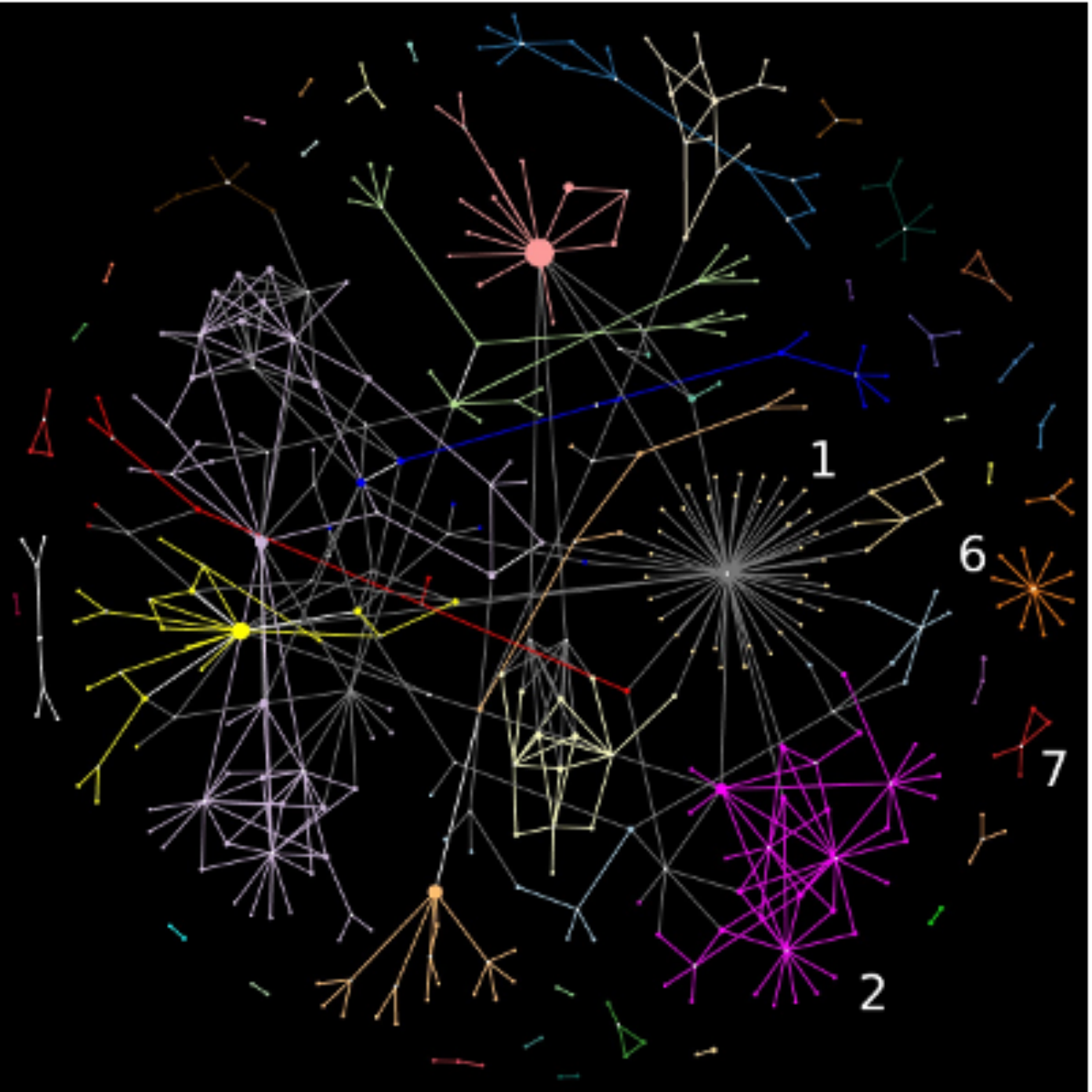}\end{center}

Sub-figure \ref{Fig exp : title:((bgo AND cryst*) OR (bgo AND calor*)) abstract:((bgo AND cryst*) OR (bgo AND calor*)) : organisations eng}
(b): extra-node representation: The coordinates of nodes are calculated
by ForceAtlas2 for this representation.

\caption{Hypergraph of organisations: Sub-figures (a) and (b) refer to the
search:\protect \\
title:((bgo AND cryst{*}) OR (bgo AND calor{*})) abstract:((bgo AND
cryst{*}) OR (bgo AND calor{*}))}
\label{Fig exp : title:((bgo AND cryst*) OR (bgo AND calor*)) abstract:((bgo AND cryst*) OR (bgo AND calor*)) : organisations eng}\selectlanguage{british}%
\end{figure}

\selectlanguage{english}%
When transfering the coordinates from the extra-node representation
to the clique view as shown on Figure \ref{Fig exp : title:((bgo AND cryst*) OR (bgo AND calor*)) abstract:((bgo AND cryst*) OR (bgo AND calor*)) : organisations eng},
the nodes tend to be well distributed on the canvas. This phenomena
finds its source in the fact that for the same hypergraph the extra-node
representation is less linked than the clique representation. Also,
when computing ForceAtlas2, the nodes tend to be more repulsed from
the centrum.

\begin{figure}[H]
\begin{center}\includegraphics[scale=0.6]{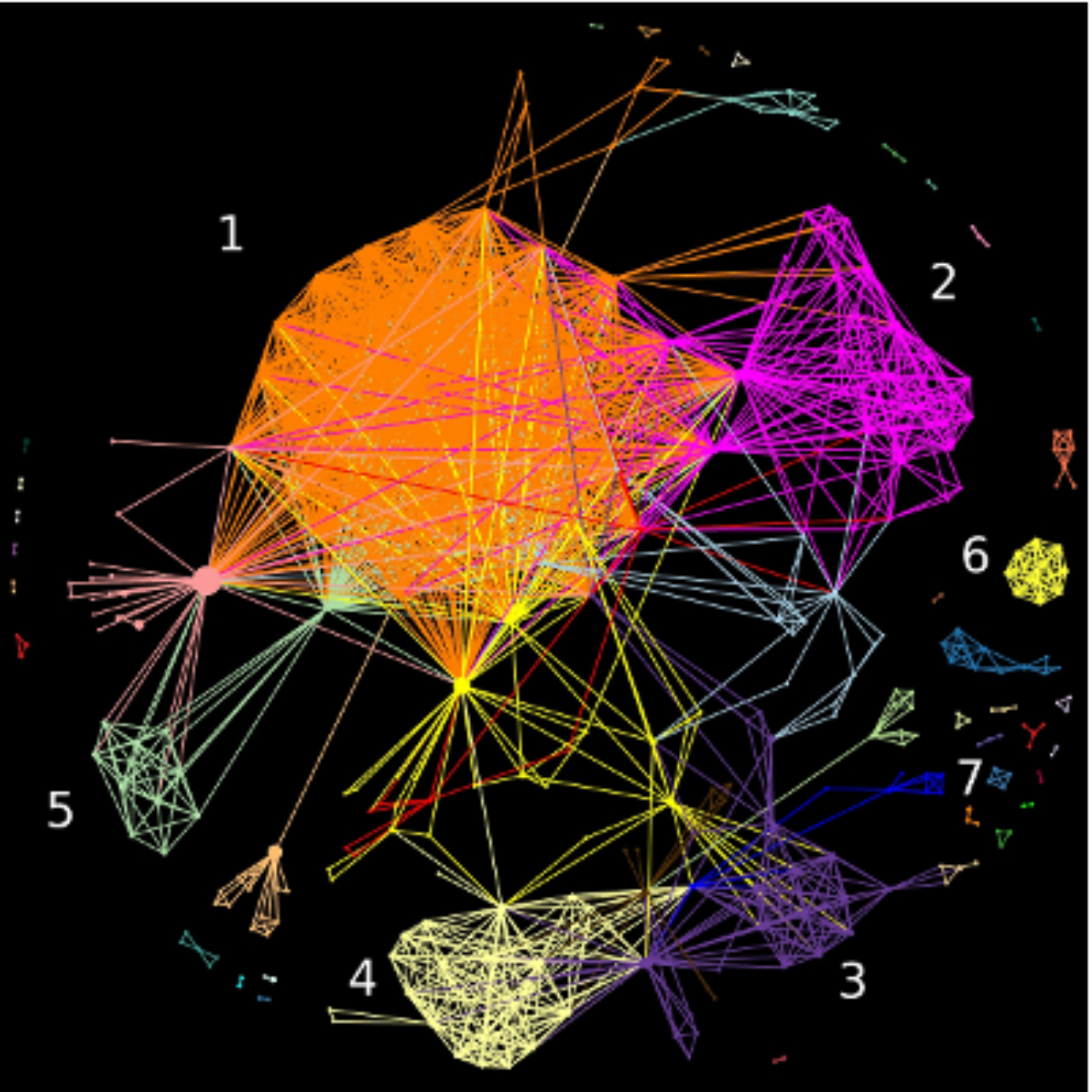}\end{center}

Sub-figure \ref{Fig exp : title:((bgo AND cryst*) OR (bgo AND calor*)) abstract:((bgo AND cryst*) OR (bgo AND calor*)) : organisations bcg}
(a): Clique representation view of the hypergraph: coordinates are
calculated by ForceAtlas2 in this representation

\begin{center}\includegraphics[scale=0.6]{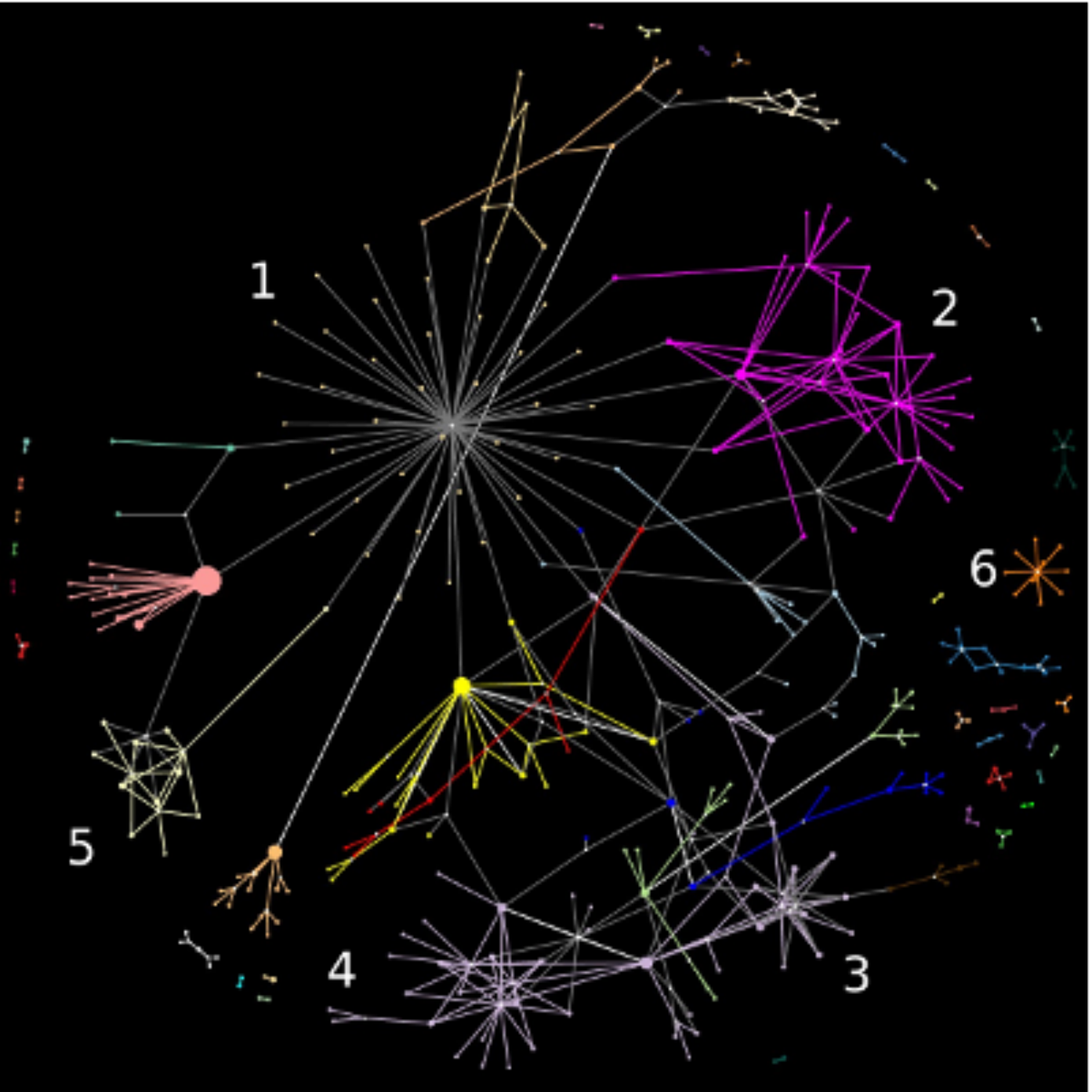}\end{center}

Sub-figure \ref{Fig exp : title:((bgo AND cryst*) OR (bgo AND calor*)) abstract:((bgo AND cryst*) OR (bgo AND calor*)) : organisations bcg}
(b): extra-node representation of the hypergraph: coordinates are
calculated by ForceAtlas2 in the clique node representation and transfered
to this one.

\caption{Hypergraph of organisations: Sub-figures (a) and (b) refer to the
search:\protect \\
title:((bgo AND cryst{*}) OR (bgo AND calor{*})) abstract:((bgo AND
cryst{*}) OR (bgo AND calor{*})): organisations}
\label{Fig exp : title:((bgo AND cryst*) OR (bgo AND calor*)) abstract:((bgo AND cryst*) OR (bgo AND calor*)) : organisations bcg}
\end{figure}

Transfering it the other way, as it is shown in Figure \ref{Fig exp : title:((bgo AND cryst*) OR (bgo AND calor*)) abstract:((bgo AND cryst*) OR (bgo AND calor*)) : organisations bcg},
lead to more gathered views, giving a better visual impact. This is
the case for all the 41 searches, both for organisations hypergraphs
and co-(author keywords) hypergraphs.

In Figure \ref{Fig exp : title:((bgo AND cryst*) OR (bgo AND calor*)) abstract:((bgo AND cryst*) OR (bgo AND calor*)) : organisations bcg},
it is interesting to see how the large collaboration numbered 1 expands
into what looks like just a one shot collaboration, while, Group 2,
which was seen as less important in the clique view appears to be
a real network of collaborators in the extra-node view. In Groups
3, 4 and 5 some internal collaborations appear that couldn't be seen
in the clique representation.

In Figure \ref{Fig exp : title:((bgo AND cryst*) OR (bgo AND calor*)) abstract:((bgo AND cryst*) OR (bgo AND calor*)) : organisations eng},
Group 6 is viewed as a single collaboration when it appears as extended
in the extra-node view, while Group 7 is viewed as one collaboration
in the clique view and in fact is represented as two collaborations
in extra-node view. 

\begin{figure}[H]
\selectlanguage{english}%
\begin{center}\includegraphics[scale=0.6]{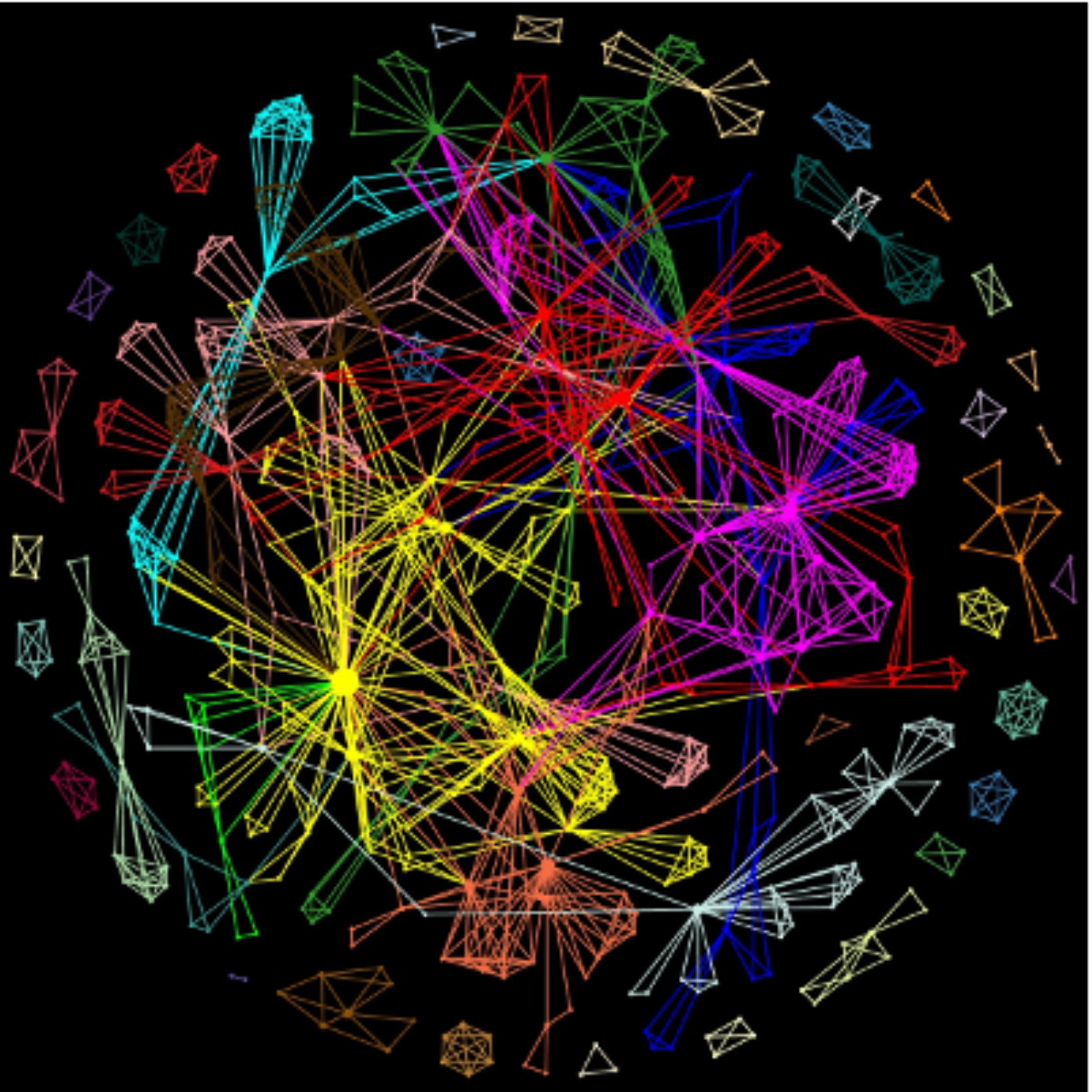}

Figure \ref{Fig exp : title:((bgo AND cryst*) OR (bgo AND calor*)) abstract:((bgo AND cryst*) OR (bgo AND calor*)) : author keywords eng}
(a) Extension of the hypergraph by clique: nodes' coordinates are
calculated on the extra-node representation.

\end{center}

\begin{center}\includegraphics[scale=0.6]{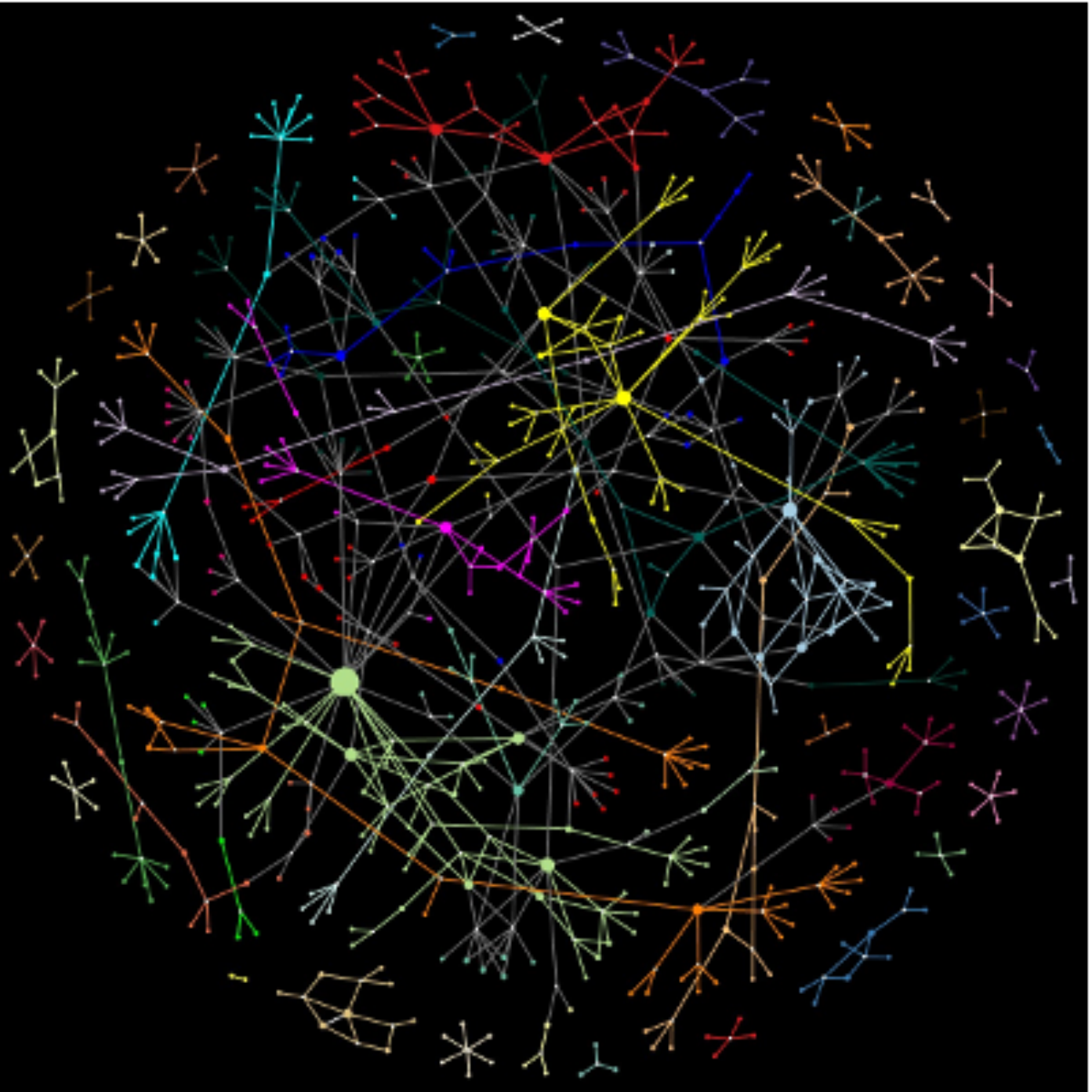}

Figure \ref{Fig exp : title:((bgo AND cryst*) OR (bgo AND calor*)) abstract:((bgo AND cryst*) OR (bgo AND calor*)) : author keywords eng}
(b) Extension of the hypergraph by extra-node: nodes' coordinates
are generated by ForceAtlas2

\end{center}

\caption{Hypergraph of author keywords: Sub-figures (a) and (b) refer to the
search:\protect \\
title:((bgo AND cryst{*}) OR (bgo AND calor{*})) abstract:((bgo AND
cryst{*}) OR (bgo AND calor{*}))}
\label{Fig exp : title:((bgo AND cryst*) OR (bgo AND calor*)) abstract:((bgo AND cryst*) OR (bgo AND calor*)) : author keywords eng}\selectlanguage{british}%
\end{figure}

The same observation can be made for co-(author keywords) as shown
in Figure \ref{Fig exp : title:((bgo AND cryst*) OR (bgo AND calor*)) abstract:((bgo AND cryst*) OR (bgo AND calor*)) : author keywords eng}
(a) and (b) where the coordinates are first calculated for the extra-node
representation of the hypergraph by ForceAtlas2 and then transfered
to the clique view. In these representations, the different peripheral
hyperedges remain the same in both views, even though the representation
is lighter in the extra-node view than in the clique view. The main
improvement is in the central part of the hypergraph for intricated
hyperedges. In this case, the extra-node view provides a good improvement
of the visualisation perception.

\begin{figure}[H]
\selectlanguage{english}%
\begin{center}\includegraphics[scale=0.6]{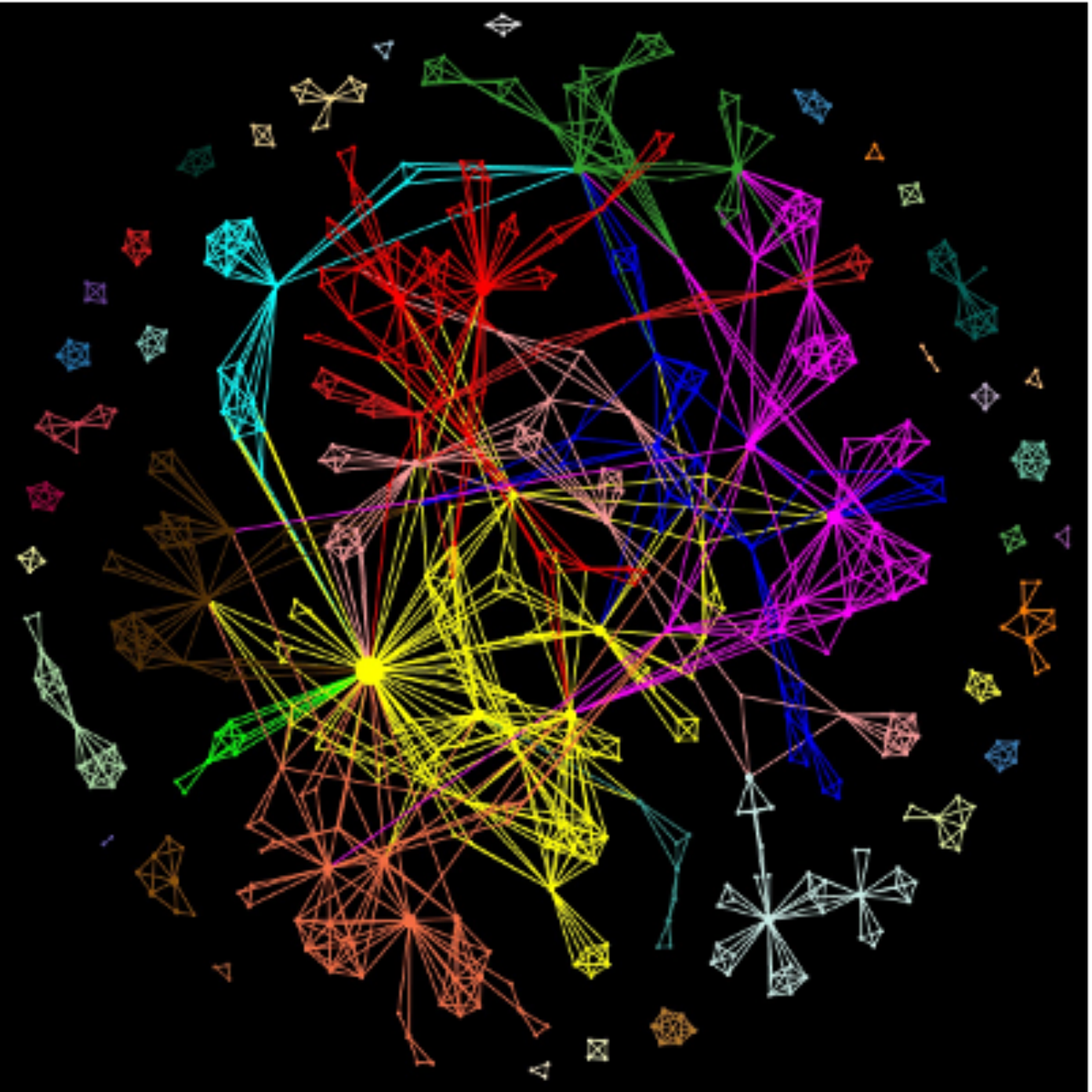}

Figure \ref{Fig exp : title:((bgo AND cryst*) OR (bgo AND calor*)) abstract:((bgo AND cryst*) OR (bgo AND calor*)) : author keywords bcg}
(a) Extension of the hypergraph by clique: coordinates are calculated
by ForceAtlas2 directly.

\end{center}

\begin{center}\includegraphics[scale=0.6]{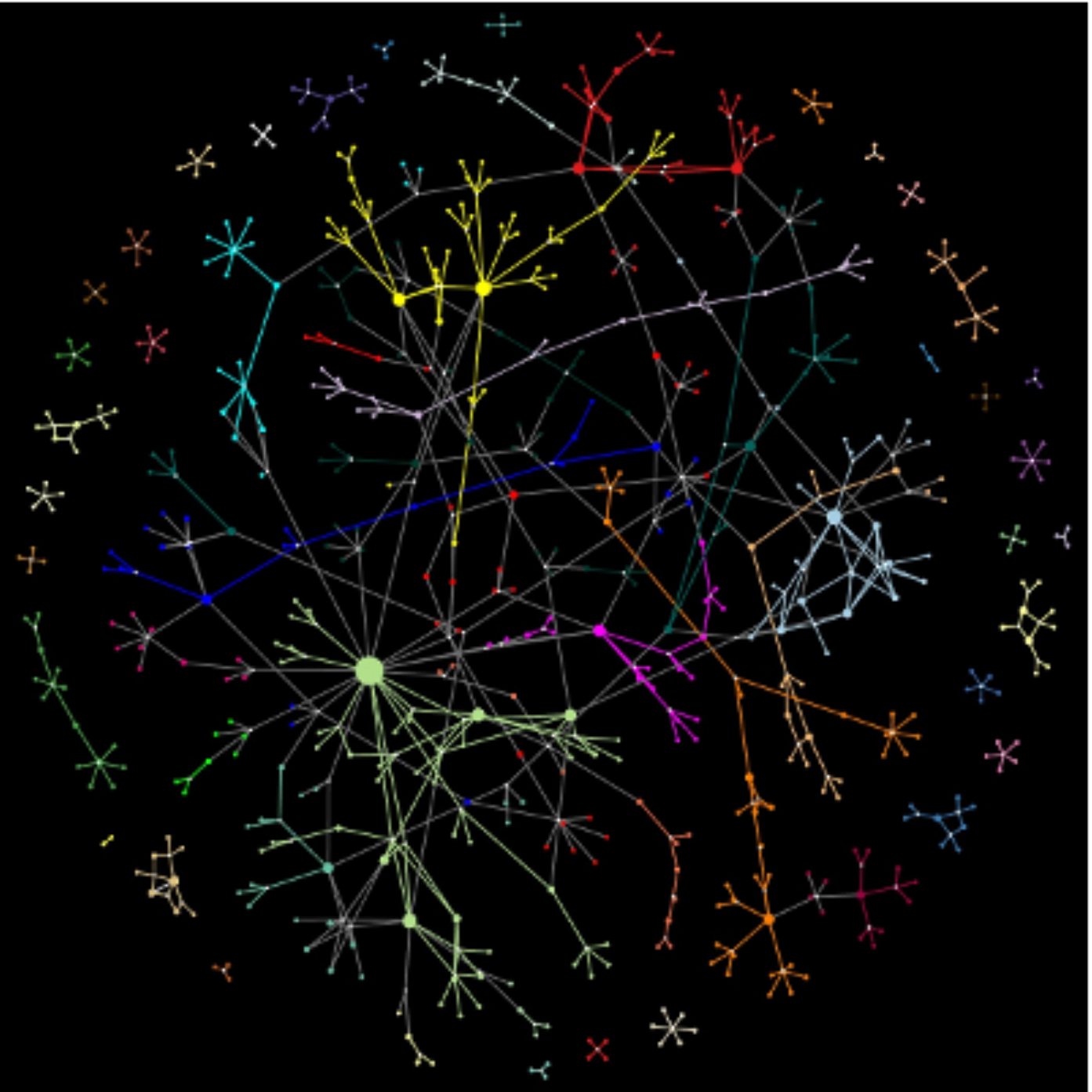}

Figure \ref{Fig exp : title:((bgo AND cryst*) OR (bgo AND calor*)) abstract:((bgo AND cryst*) OR (bgo AND calor*)) : author keywords bcg}
(b) Extension of the hypergraph by extra-node: coordinates are calculated
on the clique graph extension.

\end{center}

\caption{Hypergraph of author keywords: Sub-figures (a) and (b) refer to the
search:\protect \\
title:((bgo AND cryst{*}) OR (bgo AND calor{*})) abstract:((bgo AND
cryst{*}) OR (bgo AND calor{*}))}
\label{Fig exp : title:((bgo AND cryst*) OR (bgo AND calor*)) abstract:((bgo AND cryst*) OR (bgo AND calor*)) : author keywords bcg}\selectlanguage{british}%
\end{figure}

In Figure \ref{Fig exp : title:((bgo AND cryst*) OR (bgo AND calor*)) abstract:((bgo AND cryst*) OR (bgo AND calor*)) : author keywords bcg}
(a) and (b), the calculation of coordinates is made on the clique
view and then transferred to the extra-node view. The same remarks
than in Figure \ref{Fig exp : title:((bgo AND cryst*) OR (bgo AND calor*)) abstract:((bgo AND cryst*) OR (bgo AND calor*)) : author keywords eng}
apply. Nonetheless the computation of the coordinates by ForceAtlas2
on the clique view leads to improvements in the gathering of the nodes
when it comes to the extra-node view. Hence again, the computation
of the coordinates on the clique expansion of the hypergraph, and
their transfer to the extra-node expansion gives better results for
visualisation.

\selectlanguage{british}%

\selectlanguage{english}%
In conclusion, the visual perception is significantly enhanced in
the extra-node view. But in order to get better results, the nodes'
coordinates must be calculated taking into account the strength of
individual hyperedges, as the transfer of coordinates in between the
nodes between these two views leads to an enhanced visual perception.

In these examples the qualitative approach highlights some common
facts. The extra-node views enhance the main connecting nodes as it
is easier to visualize. The number of clusters is often greater in
the extra-node than in the clique view. This can be explained by the
smaller connectivity of the nodes. Large collaborations tend to crush
all other collaborations in the clique view while they are given their
right place in the extra-node views. For diffused hypergraphs, where
there are a lot of connections between hyperedges, the extra-node
view brings simplifications.

\subsection{Quantitative approach}

To help analysing the gain made in between the clique and extra-node
views of hypergraphs, an objectivization of the approach by quantitative
measures has been performed. The first feature that seems important
is the gain in edges. The second aims at evaluating the gain in visual
complexity of the graph. The third gives back information on the gain
in information via the calculation of entropy.

\subsubsection{Gain in edges\label{subsec:Gain-in-edges}}

Potential gain in edges on the overall dataset for organisations can
be computed. For this purpose, the size of each collaboration retrieved
from the overall dataset has been measured. This case can be seen
as the optimistic case and results has been presented in the Table
\ref{Table : Potential gain in edges organisations}, in which the
gain in edge $G_{\mbox{Edge}}$ is defined by the formula: 
\[
G_{\mbox{Edge}}=\dfrac{\left|E_{\mbox{clique}}\right|}{\left|E_{\mbox{extra-node}}\right|}
\]

where $\left|E_{\mbox{clique}}\right|$ - resp. $\left|E_{\mbox{extra node}}\right|$
- is the number of edges in the clique view - resp. extra-node view
- of the hypergraph.

Table \ref{Table : Potential gain in edges organisations} has shown
a maximal theoretical ratio that can be obtained when the hypergraph's
representation changes from clique view to extra-node view for organisations.
Nonetheless it is a maximal theoretical gain in edges as some hyperedge(s)
can be included in larger hyperedge as it has been show in Figure
\ref{Fig : Extra node unfavorable cases}. It is then of interest
to know how the gain in edges behaves in function of the average size
of collaborations in the hypergraphs.

Statistics have been performed to retrieve the real gain in edges
on each of the 63 searches. The results have been grouped by kind
of searches and the summary is presented in Table \ref{Tab : Edge ratio 63 searches}.
$Q_{1}$ is the first quartile, $Q_{2}$ the second and $Q_{3}$ the
third one.

\begin{table}[H]
\begin{centering}
\resizebox{\columnwidth}{!}{%
\begin{tabular}{|c|c|c|c|c|c|c|}
\hline 
 & $\overline{G_{\mbox{edge}}}$ & $\sigma\left(G_{\mbox{edge}}\right)$ & $Q_{1}$ & $Q_{2}$ & $Q_{3}$ & $G_{\mbox{edge}}<1$\tabularnewline
\hline 
Organisations & 4,54 & 14,27 & 0,98 & 1,13 & 1,47 & 34,9~\%\tabularnewline
\hline 
Author keywords & 1,97 & 0,29 & 1,81 & 1,97 & 2,08 & 0~\%\tabularnewline
\hline 
\end{tabular}}\newline
\par\end{centering}
\caption{Gain on edges for the 63 searches conducted}
\label{Tab : Edge ratio 63 searches}
\end{table}

\vspace{-0.5cm}

This data in Table \ref{Tab : Edge ratio 63 searches} shows that
the gain in edges is always in favour of the extra-node approach in
the case of the author keywords. For the organisations, the gain is
still in favour of the extra-node approach. The average gain is much
higher due to the fact that some searches returned very big collaborations,
exploding the number of edges in the clique view. This is due to the
high variability of size of collaboration in the case of organisations
(up to 450 collaborators, with smooth distribution) compared to author
keywords (up to 130, with more tightened distribution).

\begin{figure}[H]
\begin{center}%
\begin{tabular}{c}
\includegraphics[scale=0.35]{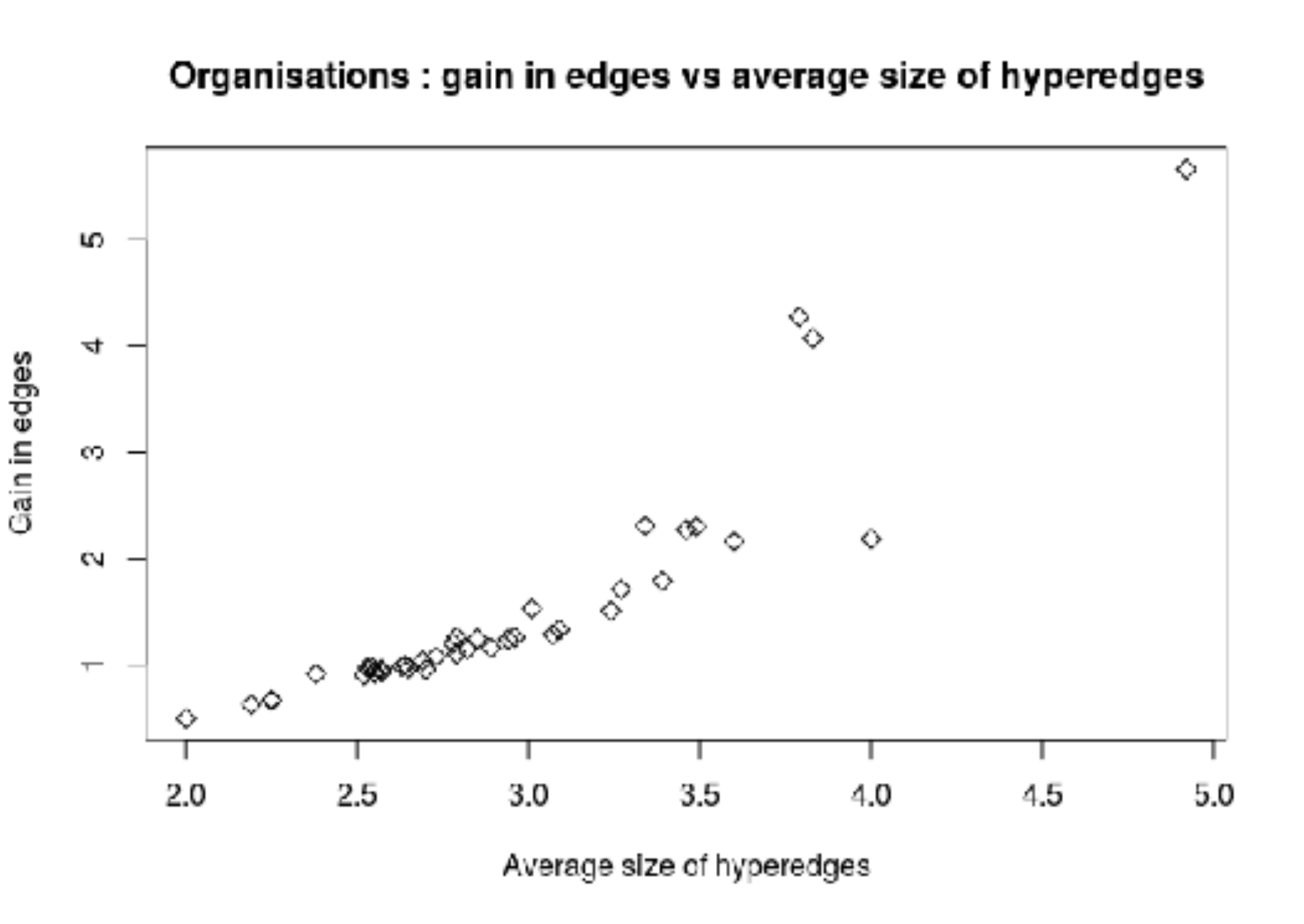}\tabularnewline
Sub-figure a: Case of the organisations\tabularnewline
\includegraphics[scale=0.35]{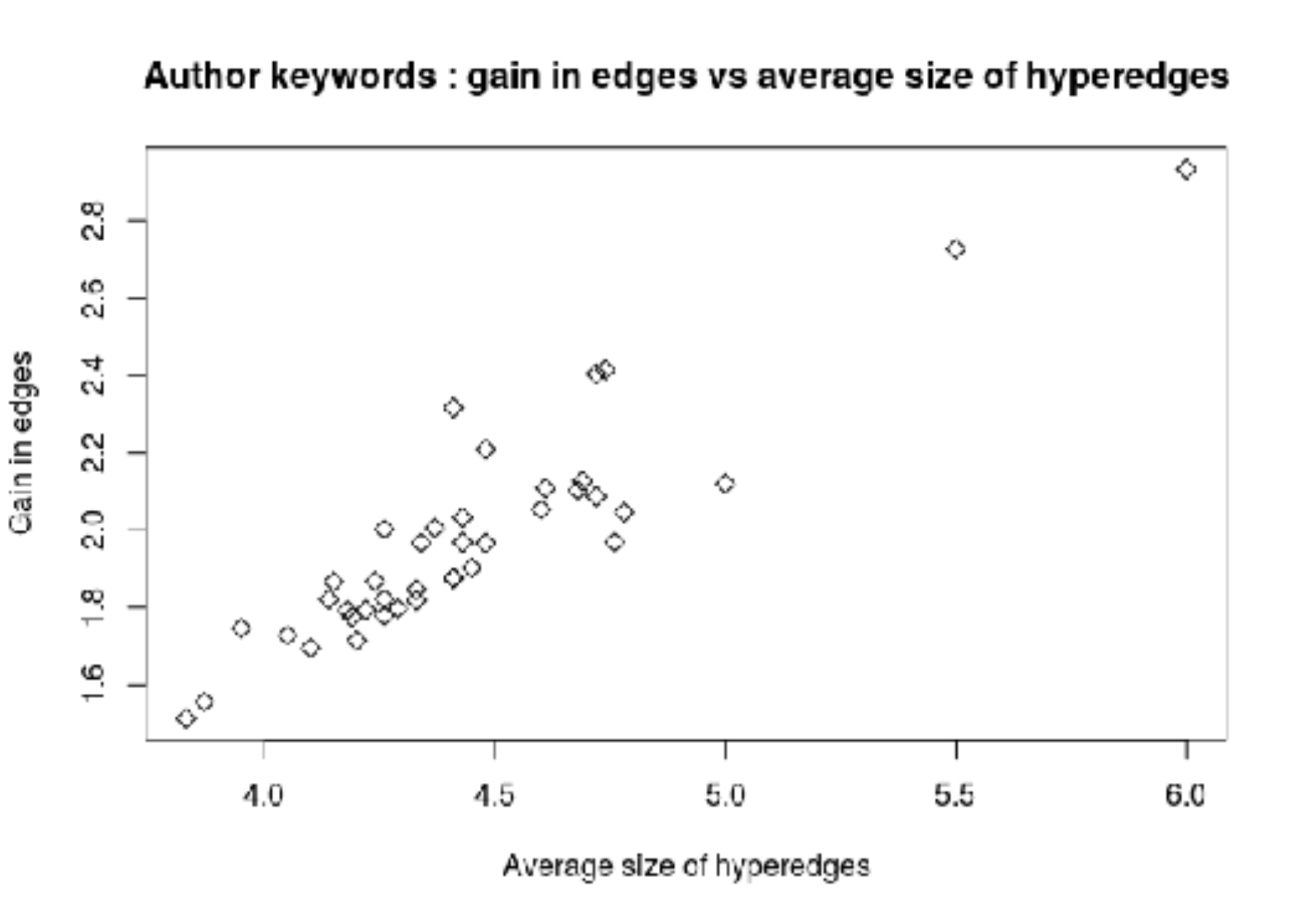}\tabularnewline
Sub-figure b: Case of author keywords\tabularnewline
\end{tabular}\end{center}

\caption{Gain on edge vs average size of hyperedges}
\label{Fig exp : Gain on edge vs average size of hyperedges}
\end{figure}

Figure \ref{Fig exp : Gain on edge vs average size of hyperedges}
shows the gain in edges versus the average size of the hyperedges.
The gain in edges increases in both cases when the average size of
hyperedges increases. It shows that the gain for organisations, with
hyperedges of average size less than 2.5 gives an increase in the
number of edges. For the author keywords, the average size of the
collaboration is bigger than in the average organisations case, and
therefore leads to a higher gain in edges for all the hypergraphs
of co-author keywords. As it can be seen on Figure \ref{Fig exp : Gain on edge vs average size of hyperedges}a,
there are some searches that are underfitting or overfitting the general
tendency due to less or more intersecting collaborations.

\subsubsection{Gain in visual complexity}

The purpose of this sub-section is to evaluate the gain in cognitive
load of the hypergraphs represented. The hypothesis made is that for
a given hypergraph where the coordinates of the nodes are similarly
calculated represented with a black background, the more black pixels
the image has, the better the contrast with the other colors will
be and the clearer the representation will be. The hypothesis on similarly
calculated is important otherwise the clearest view is a one coloured
pixel graph, which is of no interest. The clarity of a graph $G$
is introduced as 
\[
C_{G}=\dfrac{n_{\mbox{black pixels}}}{n_{\mbox{pixels in the image}}}.
\]

Also each generated hypergraph view has been exported in a raster
format, and the ratio of black pixels on the overall image has been
computed. To compare both views the clarity gain is introduced as
: 
\[
G_{C}=\dfrac{C_{\mbox{extra node view}}}{C_{\mbox{\mbox{clique view}}}}.
\]

Table \ref{Tab : Gain in clarity} shows the results for the 41 searches
that were used in the qualitative approach.

\begin{table}[H]
\resizebox{\columnwidth}{!}{

\begin{tabular}{|c|c|c|c|c|c|c|c|}
\cline{1-4} \cline{6-8} 
Force Atlas on: & \multicolumn{3}{c|}{extra-node graph} &  & \multicolumn{3}{c|}{clique graph}\tabularnewline
\cline{1-4} \cline{6-8} 
 & $C_{\mbox{clique}}$ & $C_{\mbox{extra node}}$ & $G_{C}$ &  & $C_{\mbox{clique}}$ & $C_{\mbox{extra node}}$ & $G_{C}$\tabularnewline
\cline{1-4} \cline{6-8} 
Number of images & \multicolumn{3}{c|}{41} &  & \multicolumn{3}{c|}{41}\tabularnewline
\cline{1-4} \cline{6-8} 
Average & 0.69 & 0.81 & 1.21 &  & 0.74 & 0.84 & 1.16\tabularnewline
\cline{1-4} \cline{6-8} 
Standard deviation & 0.16 & 0.11 & 0.17 &  & 0.15 & 0.10 & 0.14\tabularnewline
\cline{1-4} \cline{6-8} 
Q1 & 0.57 & 0.75 & 1.08 &  & 0.63 & 0.79 & 1.06\tabularnewline
\cline{1-4} \cline{6-8} 
Q2 & 0.73 & 0.84 & 1.16 &  & 0.78 & 0.87 & 1.12\tabularnewline
\cline{1-4} \cline{6-8} 
Q3 & 0.82 & 0.89 & 1.24 &  & 0.86 & 0.91 & 1.19\tabularnewline
\cline{1-4} \cline{6-8} 
\end{tabular}

}

\begin{center}Sub-table (a) Case of organisations\end{center}

\resizebox{\columnwidth}{!}{

\begin{tabular}{|c|c|c|c|c|c|c|c|}
\cline{1-4} \cline{6-8} 
Force Atlas on : & \multicolumn{3}{c|}{extra-node graph} &  & \multicolumn{3}{c|}{clique graph}\tabularnewline
\cline{1-4} \cline{6-8} 
 & $C_{\mbox{clique}}$ & $C_{\mbox{extra node}}$ & $G_{C}$ &  & $C_{\mbox{clique}}$ & $C_{\mbox{extra node}}$ & $G_{C}$\tabularnewline
\cline{1-4} \cline{6-8} 
Number of images & \multicolumn{3}{c|}{41} &  & \multicolumn{3}{c|}{41}\tabularnewline
\cline{1-4} \cline{6-8} 
Average & 0.54 & 0.71 & 1.35 &  & 0.64 & 0.78 & 1.25\tabularnewline
\cline{1-4} \cline{6-8} 
Standard deviation & 0.12 & 0.11 & 0.13 &  & 0.12 & 0.09 & 0.11\tabularnewline
\cline{1-4} \cline{6-8} 
Q1 & 0.44 & 0.65 & 1.28 &  & 0.55 & 0.73 & 1.17\tabularnewline
\cline{1-4} \cline{6-8} 
Q2 & 0.53 & 0.73 & 1.37 &  & 0.65 & 0.81 & 1.24\tabularnewline
\cline{1-4} \cline{6-8} 
Q3 & 0.60 & 0.77 & 1.43 &  & 0.72 & 0.83 & 1.31\tabularnewline
\cline{1-4} \cline{6-8} 
\end{tabular}

}

\begin{center}Sub-table (b) Case of author keywords\end{center}

\caption{Clarities and gain in clarity}
\label{Tab : Gain in clarity}
\end{table}

\vspace{-0.5cm}

For organisations, the average clarity of the extra-node view is always
better. The gain is better in extra-node coordinates' computation,
but the clarity remains lower than the one obtained in the clique
coordinates' computation in the extra-node view. This is the case
for all the 41 searches in accordance with what was expected in the
qualitative approach.

Similar results are obtained for author keywords; the best clarity
is obtained for the extra-node view where coordinates have been calculated
on applying ForceAtlas2 to the clique view and transfered to the extra-node
view.

As a conclusion, this gain index confirms that the best clarity is
obtained with the extra-node view, independently of the way of calculating
the coordinates in most of the case for organisations and always for
author keywords. 

\subsubsection{Entropy and gain in information}

In this section the entropy of the images will be calculated to show
that graphs generated with ForceAtlas2 on clique view and transfer
to extra-node view are better organized. 

The entropy is a good way to know the degree of organisation of an
image. If a set of things is well organized then the entropy is low.
On the other end, if things are not organized entropy is high. Entropy
was first introduced by \cite{SHANNON 1948}. 

The entropy is defined here as : 
\[
H=-\left(C_{G}\log_{2}\left(C_{G}\right)+\left(1-C_{G}\right)\log_{2}\left(1-C_{G}\right)\right).
\]

A uniform distribution of black pixels achieved when $C_{G}=0.5$
maximizes the entropy to a value of 1. The lower the entropy the more
organized the rendering will appeared.

Table \ref{Tab : Gain in entropy} shows the results for the 41 searches
that were used in the qualitative approach.

\begin{table}[H]
\resizebox{\columnwidth}{!}{

\begin{tabular}{|c|c|c|c|c|c|}
\cline{1-3} \cline{5-6} 
Force Atlas on: & \multicolumn{2}{c|}{extra-node graph} &  & \multicolumn{2}{c|}{clique graph}\tabularnewline
\cline{1-3} \cline{5-6} 
 & $H_{\mbox{clique}}$ & $H_{\mbox{extra node}}$ &  & $H_{\mbox{clique}}$ & $H_{\mbox{extra node}}$\tabularnewline
\cline{1-3} \cline{5-6} 
Number of images & \multicolumn{2}{c|}{41} &  & \multicolumn{2}{c|}{41}\tabularnewline
\cline{1-3} \cline{5-6} 
Average & 0.80 & 0.65 &  & 0.75 & 0.60\tabularnewline
\cline{1-3} \cline{5-6} 
Standard deviation & 0.19 & 0.20 &  & 0.20 & 0.20\tabularnewline
\cline{1-3} \cline{5-6} 
Q1 & 0.68 & 0.51 &  & 0.60 & 0.44\tabularnewline
\cline{1-3} \cline{5-6} 
Q2 & 0.85 & 0.64 &  & 0.77 & 0.55\tabularnewline
\cline{1-3} \cline{5-6} 
Q3 & 0.96 & 0.80 &  & 0.95 & 0.73\tabularnewline
\cline{1-3} \cline{5-6} 
\end{tabular}

}

\begin{center}Sub-table (a) Case of organisations\end{center}

\resizebox{\columnwidth}{!}{

\begin{tabular}{|c|c|c|c|c|c|}
\cline{1-3} \cline{5-6} 
Force Atlas on : & \multicolumn{2}{c|}{extra-node graph} &  & \multicolumn{2}{c|}{clique graph}\tabularnewline
\cline{1-3} \cline{5-6} 
 & $H_{\mbox{clique}}$ & $H_{\mbox{extra node}}$ &  & $H_{\mbox{clique}}$ & $H_{\mbox{extra node}}$\tabularnewline
\cline{1-3} \cline{5-6} 
Number of images & \multicolumn{2}{c|}{41} &  & \multicolumn{2}{c|}{41}\tabularnewline
\cline{1-3} \cline{5-6} 
Average & 0.95 & 0.82 &  & 0.90 & 0.71\tabularnewline
\cline{1-3} \cline{5-6} 
Standard deviation & 0.08 & 0.16 &  & 0.11 & 0.17\tabularnewline
\cline{1-3} \cline{5-6} 
Q1 & 0.937 & 0.772 &  & 0.858 & 0.660\tabularnewline
\cline{1-3} \cline{5-6} 
Q2 & 0.980 & 0.837 &  & 0.934 & 0.703\tabularnewline
\cline{1-3} \cline{5-6} 
Q3 & 0.996 & 0.935 &  & 0.990 & 0.843\tabularnewline
\cline{1-3} \cline{5-6} 
\end{tabular}

}

\begin{center}Sub-table (b) Case of author keywords\end{center}

\caption{Entropies}
\label{Tab : Gain in entropy}
\end{table}

\vspace{-0.5cm}

These results show in both cases - author keywords and organisations
- that the entropy is the lowest in the extra-node view with coordinates
calculated by the clique representation. It confirms that this representation
of hypergraphs gives the best results in term of structured information.

\subsubsection{Quantitative approach main teachings}

Two indices have been built to help quantifying the gain both in visuality
and in the number of edges. The clarity index is particularly relevant
to quantify the quality of the final view of the hypergraph. This
is confirmed by the entropy indicating how well structured the views
are.

On the one hand this clarity index is used in comparison between the
two views of the same hypergraph, and it is more the relative positioning
and gain than the absolute value of this index that is important.
On the other hand the entropy allows a global comparison.

Both clarity index and entropy show that the best approach for the
visualisation for author keywords hypergraphs is nearly always in
the extra-node view with coordinates calculated in ForceAtlas2 in
clique view. Entropy confirms with more strength the results obtained
by clarity by itself.

The gain in edges when switching from the clique view to the extra-node
view is also often bigger, and in very large graphs it can be quite
high. This can solve some problems of computability for large graphs
with very large collaborations.

\section{Conclusion}

Hypergraphs allow a better rendering of the structure of publications,
and retain in some kind the footprint of the article structure.

Hypergraphs rendering can bring a lot of visual information. The visual
rendering of such hypergraphs is challenging, and was central to this
research. The clique view approach allows the rendering of large hyperedges,
but hides small colaborations in between players of large collaborations.
These large collaborations are over emphasized even when occuring
only once. This phenomena disappears in the extra-node view where
each collaboration has the same visual impact potential, and can be
viewed directly.

Nonetheless, the tied links of hyperedges which are expressed in the
clique view allows the gathering of the nodes when it comes to place
them. In the extra-node view, as the nodes are less linked, they tend
to spread out all over the view. Therefore this study provides a way
of setting coordinates to the nodes of the extra-node view taking
into account hyperedges' specificity through the calculation of coordinates
via the clique view. Moreover the experimental part brings a positive
answer to the usage of the extra-node view for visualisation of hypergraphs
of collaborations. Using organisations and author keywords were the
distribution of the cardinalities of hyperedges are very different
allows to generalize this result. Furthermore this work highlights
new indicators that can help evaluating the visual impact and gain
in cognitive load of the chosen representation.

This work has shown that the extra-node view is a reliable way of
showing hypergraphs where hyperedges' structure is preserved. Hypergraphs
are a useful model for collaborative network and their visualisation.
Future work should include a study on other nodes' placements using
other rendering than the one of ForceAtlas2 to confirm the results
obtained in this paper. This can be done by focusing on new layouts
of hypergraphs that can enhanced the visual perception of the data
set and the enrichment allowed by hypergraphs.

\section{Acknowledgments}

We are really thankful to all the team of Collaboration Spotting from
CERN - supervised by Jean-Marie LE GOFF - for all the exchanges we
had: Adam AGOCS, Dimitris DARDANIS, Dimitri PROIOS, and Tim HERWERCK.

This work was started during the master thesis of Xavier Ouvrard,
who is really thankful to Laurent PHILIPPE of University of Bourgogne
Franche-Comté for his supervision. This work is continued during the
PhD of Xavier OUVRARD, done at CERN in the Collaboration Spotting
project.\\
\\
\\
\\
\\
\\

\end{document}